\documentstyle[12pt,epsfig]{article}
\begin{document}


\vskip 1cm

\title{\bf 
Slow nonequilibrium dynamics: parallels between 
classical and quantum glasses and gently driven systems$\;$
\footnote{Contribution to the 4th Discussion Meeting
on Complex phenomena, Creta, June 2001}
}
\vskip 20pt
\author{
{\bf  Leticia F. Cugliandolo\footnote{Research associate
ICTP-Trieste, Italia.}}
\\
 Laboratoire de Physique Th{\'e}orique, Ecole Normale Sup{\'e}rieure,
\\
24 rue Lhomond 75231 Paris Cedex 05 France and 
\\
Laboratoire de Physique Th{\'e}orique  et Hautes {\'E}nergies, Jussieu, \\
1er {\'e}tage,  Tour 16, 4 Place Jussieu, 75252 Paris Cedex 05, France
}

\date\today
\maketitle
\begin{abstract}
We review an scenario for the
non-equilibrium dynamics of glassy systems that
has been motivated by the exact solution of simple models.
This approach allows one to set on firmer grounds
well-known phenomenological theories. The old ideas of
entropy crisis, fictive temperatures, free-volume...
have clear definitions within these models.
Aging effects in the glass phase are also captured.
One of the salient features of the analytic solution, 
the breakdown of the fluctuation-dissipation
relations, provides a definition of a bonafide
{\it effective temperature} that is measurable by a
thermometer, controls heat flows, partial equilibrations,
and the reaction to the external injection of heat.
The effective temperature is an extremely robust
concept that
appears in non-equilibrium systems in the
limit of small entropy production as, for instance,
sheared fluids, glasses at low temperatures when quantum fluctuations
are relevant, tapped or vibrated granular matter, etc.
The emerging scenario is one of partial equilibrations,
in which glassy systems arrange their internal degrees of freedom
so that the slow ones select their own effective temperatures.
It has been proven to be consistent within any perturbative
resummation scheme (mode coupling, etc) and
it can be challenged by experimental and numerical
tests, some of which it has already passed.
\end{abstract}

\newpage

\section{Introduction}

The theoretical understanding of the 
non-equilibrium dynamics of glassy systems is now being reached.
Setting the accent on the dynamics, and leaving 
aside subtle and possibly undecidable questions as whether there exists a 
thermodynamic glass transition, allows for a unified analytical 
treatment of many glassy phenomena. Indeed, in spite of important 
differences in the microscopic constituents, the interactions among them,
and the nature of the microscopic
dynamics, many of the macroscopic dynamic features 
of glassy materials are similar and can  
be captured by a generic approach.  
It is the aim of this article to review the scenario for the 
out of equilibrium dynamics of glassy systems 
that emerges from the study of simple test models~\cite{review}. 
The outcome of 
its detailed analysis can be, and to a certain extent has been, put to test 
by comparing it to the numerical analysis of more realistic models 
and to experiments in real systems.

Glassy states of matter are ubiquitous in nature. When a liquid is 
annealed at a sufficiently rapid rate, it reaches a temperature region 
in which its dynamics can no longer follow the pace imposed by the thermal 
environment. The system enters the glassy phase where it cannot equilibrate.
The relaxation at all lower temperatures 
occurs out of equilibrium. One of its hallmarks are 
aging effects, or the impossibility of reaching stationarity~\cite{aging}. 
Typical examples of
such glassy systems are polymer glasses, spin glasses, structural 
glasses, orientational glasses, vortex glasses, etc. In all these systems the 
crossover temperature is sufficiently high as to make quantum effects 
totally irrelevant.     

Quantum glassy phases, where quantum fluctuations are at least 
as important as thermal ones, 
 have been identified in a number of materials.
Two such examples are the spin glass compound
and the amorphous insulator studied in  
\cite{Aeppli} and \cite{Osheroff}, respectively. 
Another interesting realization is the so-called
Coulomb glass in which localized electrons interact via Coulomb 
two-body potentials and hop between localization centers~\cite{Zvi}. 
In both systems the dynamics is extremely slow and  
strong history dependences as 
well as other glassy features have been observed~\cite{Aeppli,Osheroff,Zvi}.

The above examples concern systems that are not able to reach 
equilibrium with its environment in measurable times but that, let
evolve on astronomical time-scales, will eventually equilibrate.
Other ways of establishing non-equilibrium states with slow dynamics 
are also possible. For instance, a dense liquid can be driven to a slow 
out of equilibrium stationary regime by a weak shear. This perturbation 
accelerates the dynamics in such a way that the structural relaxation 
time (equivalently the viscosity) decreases with increasing shear strength 
(shear thinning)~\cite{rheology}.  
Non-potential forces also have a strong effect on aging systems:
they introduce a time scale beyond which aging is 
interrupted.
  
The driven dynamics of granular matter is now receiving renewed 
attention~\cite{review_sid}.
Since the potential energy needed to displace a macroscopic
grain by a distance equal to 
its diameter is much larger than the characteristic thermal energy, $k_BT$,
thermal activation is totally irrelevant for granular systems.
Therefore, granular systems are blocked in 
metastable states in the absence of external driving. 
Instead, when energy is pumped in in the form of 
shearing, vibration or tapping, transitions between the otherwise 
stable states occur and granular matter slowly relaxes toward 
configurations with higher densities.
Glassy features such as hysteresis
as a function of the amount of energy injected, slow dynamics, and 
non stationary correlations have been exhibited.  

Many of the recent  developments in the understanding of the similarities 
in the behavior of a priori so different systems  
are based on the analysis of simple mean-field-like models.
In a series of seminal papers Kirkpatrick, Thirumalai and 
Wolynes (KTW)~\cite{Kithwo} showed that 
disordered spin models such as a fully connected Ising model 
with random $p$ spin interactions ($p\geq 3$), or the Potts glass, 
capture many of the 
expected properties of  the glass dynamic transition. Consequently, 
it realizes exactly some of the older phenomenological descriptions. 
Neither the fact that the dynamic variables are spins
nor the presence of disorder
are essential; they are used only as a calculational tool. 
This was signaled by these authors 
and later made explicit in a series of articles
 where non disordered models with similar phenomenology
have been proposed and studied~\cite{review}.

In short, KTW showed the following. In the liquid phase
the {\it exact} dynamic equation for the auto-correlation 
function  coincides with
the schematic mode-coupling equation~\cite{Gotze}.  
Hence, the dynamics of these models in the high temperature phase 
captures the same physics as the schematic mode-coupling theory and, in 
particular,  the dynamic transition at $T_d$ ($\equiv T_{\sc mct}$) 
toward a glassy phase. The dynamic transition is {\it discontinuous}, 
in the sense that a plateau in the decay to zero of the correlations
already appears in the (equilibrium) liquid phase when $T_d$ is 
approached from above; it is of second order in the usual 
sense since thermodynamic quantities as, for instance, the 
energy density are continuous when going across $T_d$.    
The dynamic transition is due to the 
proliferation of metastable states as proved by the analysis of the
Thouless-Anderson-Palmer ({\sc tap}) free energy density. 

Since the model is defined by a simple Hamiltonian, the static 
properties are also easily accessible.
A static transition between a liquid and a glassy thermodynamic phase
occurs at a lower temperature $T_s$ ($<T_d$) where the configurational 
entropy vanishes. This phenomenon is related to the Kauzmann paradox;
$T_s$ is then associated to $T_K$. The static transition is also 
discontinuous since the order parameter, $q_{\sc ea}$,  jumps at $T_s$ 
but it is of second order thermodynamically. 

The relaxational dynamics below $T_d$ was solved in \cite{Cuku}. The 
model does not reach equilibrium with the thermal bath unless exponentially 
diverging times with $N$, the number of degrees of freedom, are 
considered. The non-equilibrium solution is then relevant for most 
cases of practical interest, since such diverging times
are unrealistic. One of the main features of the analytic solution
is that aging effects, much as those observed experimentally, 
are captured by this model.
The reason why they subsist asymptotically 
can be grasped from the analysis of the metastable states, 
again using {\sc tap}'s approach. Extensions of the $p$ spin-glass model 
to other mean-field models with internal dimension
lead to predictions about spatio-temporal scalings~\cite{space}.
Above $T_d$ these are equivalent to the extensions of the mode-coupling 
approach to super-cooled liquids~\cite{Gotze}; below $T_d$ they are 
related to mode-coupling approximations of the realistic (hard to treat) 
models in which the assumption of equilibration is not 
done~\cite{Mode,Arnulf}. 

It is important to stress that, even if the 
$p$ spin model has quenched disorder in its interactions, it is not 
a spin-glass model. It is rather well established that spin-glasses (in the absence of an 
external field) have second order phase transitions with no discontinuity
of any order parameter. Moreover $T_s=T_d$ for a spin-glass. 

In the models and approximations mentioned above, 
the dynamic and static transitions are sharp.  
In real glasses, $T_d$ is a crossover while $T_s$
might not exist. In spite of this and other defects, the interest 
of these solutions is that they have a great power of prediction
of so far unknown effects.
The rather accurate comparison of these to numerical 
simulations~\cite{numerics} and, to the extent of 
their availability, experiments~\cite{experiments,aging_2,Teff_exp}, 
supports the proposal
that the  mechanism in these models is similar to the one 
responsible for the glass transition, and the glassy dynamics,
in real materials. Moreover, some of the ingredients missing
in the full analytic solution of the mean-field models that will
render their description of real materials more accurate,
have been identified (preasymptotic corrections when the thermodynamic 
limit has been taken, analysis of the dynamics in time-scales 
that diverge with $N$, the r\^ole played by fluctuations, etc.). 
For the moment, their analytical 
treatment has proven too difficult.

Another important consequence of the study of these solvable models is that 
it has allowed us to identify an
 effective temperature, $T_{\sc eff}$, that can be measured experimentally
in a direct manner~\cite{Cukupe}. This quantity has 
all the expected properties of a temperature: 
it controls heat transfer, partial 
equilibrations, and the reaction to external heat transfers. 
More importantly, different observables that evolve in the same 
time-scale and interact should have the same effective temperature.  
The solution of the models and approximations with 
space and time~\cite{space,Arnulf} do indeed respect this property. 
Recent numerical simulations in Lennard-Jones mixtures have also
shown the validity of this important property~\cite{Ludo-unpub}.
This concept lead to rather straightforward
extensions of the dynamic scenario to other non-equilibrium systems
with small entropy production as  quantum 
glasses~\cite{Culo,quantum-others,Bicu}, 
vibrated glasses~\cite{jorge,granular_us} 
and weakly perturbed liquids and 
glasses~\cite{jorge,Cukulepe,rheology_theor}.  
In our opinion, the mean-field models are explicit and 
solvable realizations of a general scenario~\cite{Cuku-japan}  
based on the generation of effective temperatures 
in generic non-equilibrium systems with slow dynamics. 

In this respect, it is interesting to compare the effective temperature
to the proposal of 
Tool, Narayanawasmy, Moynihan and collaborators~\cite{fictive} to
introduce phenomenological {\it fictive temperatures}
in order to characterize the departure from equilibrium of 
glasses.  It is important to stress that, even if similar in 
spirit, the effective temperature is more than a phenomenological
definition. In particular, the fictive temperature is defined 
with respect to a given physical property and it can 
vary from one property to another. The effective temperature 
instead has to be the same for all physical properties
(that interact) when studied in the same time-scales.

The definition of a bonafide effective temperature has rekindled
the search for an extension of thermodynamics to describe the quasi-static 
properties of the glassy non-equilibrium dynamics.
In fact, Moynihan {\it et al} had studied the modification of 
several thermodynamic relations, {\it e.g.} the Ehrenfest relations and 
the Prigogine-Defay ratio,  when a (time-dependent) fictive temperature is
added as an ``internal'' parameter~\cite{Moynihan2}. 
More recently, these questions have been revisited~\cite{Ni}
using the effective temperature as a supplementary parameter
in the thermodynamic potentials.
This was also motivated by the solution to $p$ spin-like models for 
which ``mixed'' thermodynamic potentials are relevant.

Another interesting problem is the 
explicit search of a relation between the effective 
temperature and the structure 
of metastable states in the free-energy landscape. 
Again, the exact results for the mean-field 
models acted as a guideline of immense help. Indeed, for $p$ spin like 
models \cite{Remi,jorge} the value of $T_{\sc eff}$
 coincides with a quasi-static quantity
defined as $\partial S^{\sc tap}(f,T)/\partial f|_{f_{\sc th}}$ with
$f$ the {\sc tap} free-energy density, 
$S^{\sc tap}$ the logarithm of the number of stationary points of
 the {\sc tap} free-energy density
in the interval $[f,f+df]$, and $f_{\sc th}$ a particular 
value with a dynamical meaning (it is the 
free-energy of the {\it threshold} level, the one related to the energy
density reached asymptotically~\cite{Cuku,Bi}). This result has
been subsequently analysed in more general terms in \cite{Frvi}.

The relation between the effective temperature and 
the {\sc tap} entropy makes contact with the proposal
of Edwards~\cite{Edwards} that an ensemble of {\it equiprobable 
blocked} configurations describes the statistical properties of 
gently driven granular matter in the steady state. (It is important 
to stress that the definition of blocked configuration actually depends 
on the dynamics, {\it e.g.} in a spin system blocked may mean 
stable with respect to single spin flips.) Edwards'
entropy, $S_{\sc e}({\cal E})$, is defined as the logarithm of the number 
of blocked configurations with energy density in $[{\cal E}, 
{\cal E}+d{\cal E}]$, and Edwards' temperature is then
$T_{\sc e}^{-1} = \partial S_{\sc e}({\cal E})/\partial {\cal E}$.
It has been recently
checked that this definition coincides with the effective temperature
in a number of glassy models at zero
temperature~\cite{Edwards-check,Makse}; some 
cases where it does not work  have also been exhibited, showing that the 
connection cannot be generically valid. An early experimental attempt to 
test the assumption of a flat distribution in a granular system
has been performed by Nowak {\it et al}~\cite{Nowak}. 
 
Similar ideas have been 
explored~\cite{Sciortino,Crisanti, Cavagna} in the context of 
the inherent structure 
construction of Stillinger and Weber~\cite{Stillinger},  and its
extensions. Mainly, the idea is to relate the 
aging dynamics of, {\it e.g.}, a finite size mean-field model~\cite{Crisanti}
or a Lennard-Jones glass former~\cite{Cavagna,Sciortino}
to the evolution  in the multidimensional
{\it potential energy landscape} once partitioned in basins 
of minima and saddles.
Thermal activation is totally irrelevant in granular matter 
and the use of energy densities, instead of free-energy densities, is 
justified in this case. At finite
temperature potential energy densities and free-energy densities 
are not equivalent. Hence, the connection between the approaches based on the 
study of the potential energy landscape and 
the ones based on mean-field models at finite temperatures have to be 
taken with care~\cite{Bimo}.

\section{Slow relaxation in systems out of equilibrium}

After having very briefly described the main ingredients of the 
behaviour of freely relaxing mean-field like models, 
let us now compare  in more detail
the main features of the dynamic relaxation in 
free evolving classical and quantum glassy systems, and externally
driven complex liquids and glasses.

We have argued that the experimentally relevant sizes and times 
are such that the thermodynamic limit has to be taken
first and the asymptotic limit can be taken only afterwards.
When a transient following preparation has elapsed, 
in all the systems and experimental setups described in the introduction
(with the proviso that energy has to be weakly injected in the case of driven 
systems) the macroscopic relaxation occurs in  several 
time scales. The separation of time scales is most clearly shown 
by the evolution of the self-correlation, in the case of spin models, 
or by the decay of the incoherent scattering function in the case of 
particle systems. We denote both by $C_k(t+t_w,t_w)$, with $k$ a chosen 
wave-vector (usually, $k=0$ for the spin model), $t_w$ the waiting-time
measured from the moment in which the experimental 
glass transition $T_g$ is crossed and $t+t_w \geq 
t_w$ the measuring
time. 

Above the glass transition, super-cooled liquids have a stationary relaxation
for all waiting times that are longer than the finite equilibration time.
$C_k(t)$ has a rich behavior with a rapid decay toward a plateau 
(visible in logarithmic scale) and a second slow decay toward zero that 
occurs in the $\alpha$ relaxation time $t_\alpha(T)$. 

\begin{figure}[h]
\centerline{
\epsfxsize=3in
\epsffile{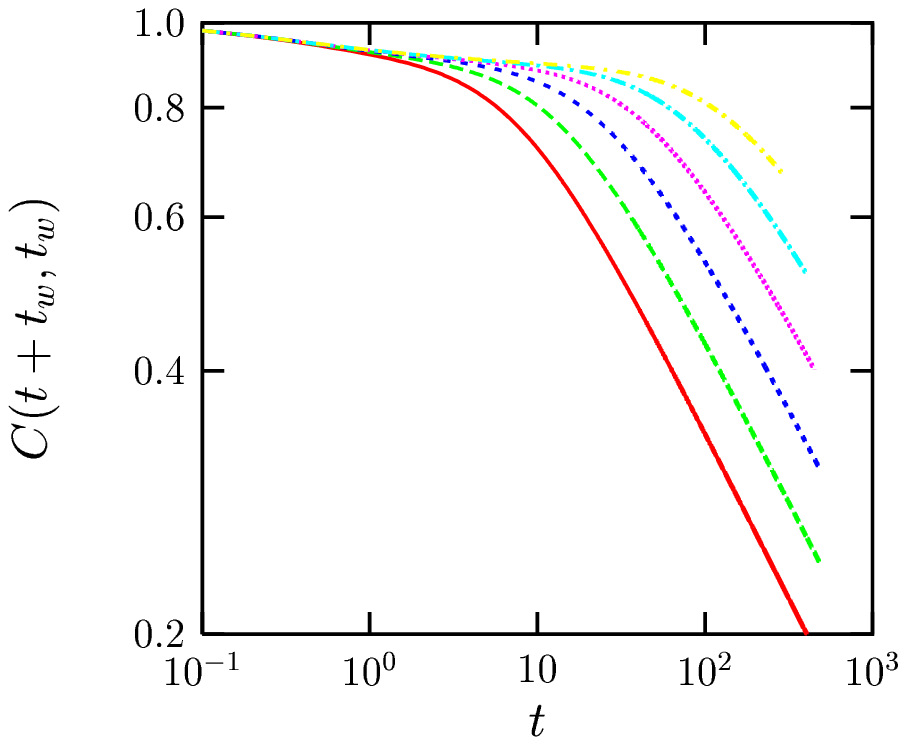}
\epsfxsize=3in
\epsffile{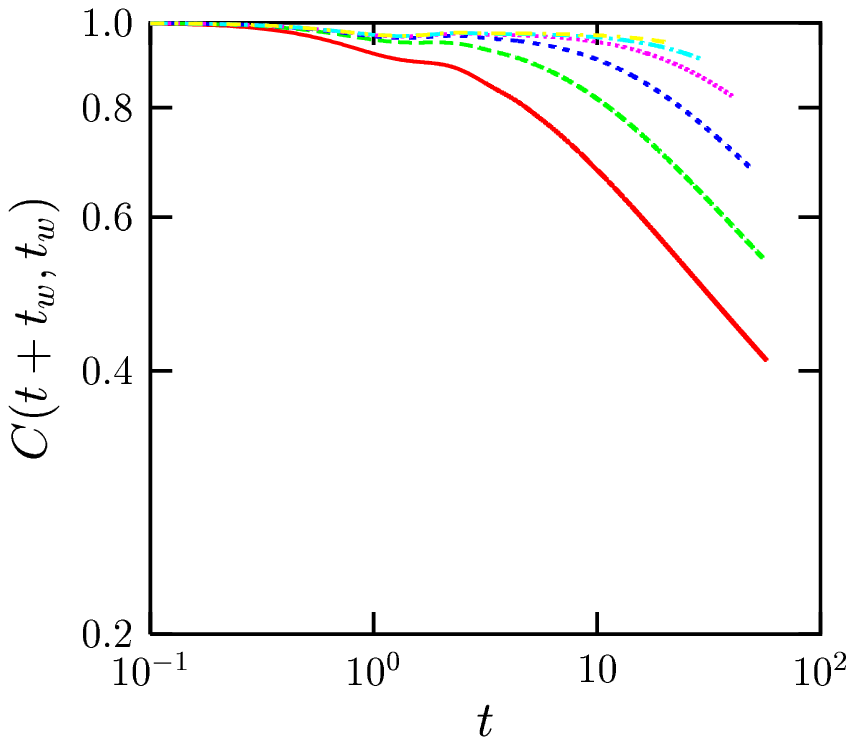}
}
\centerline{
\hspace{.7cm}
\epsfxsize=3in
\epsffile{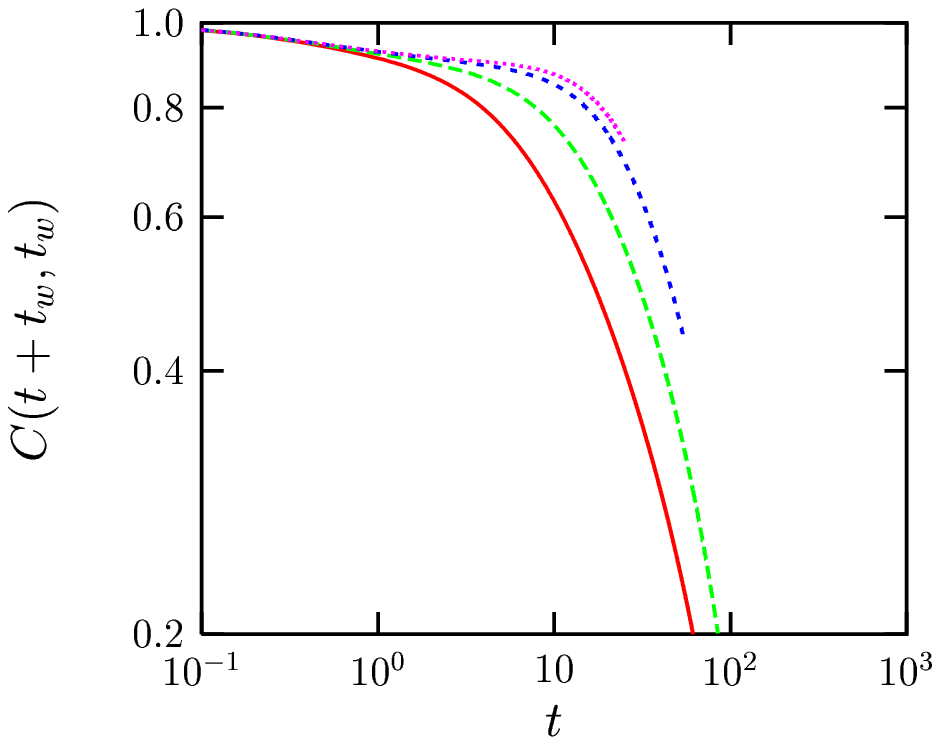}
\epsfxsize=3in
\epsffile{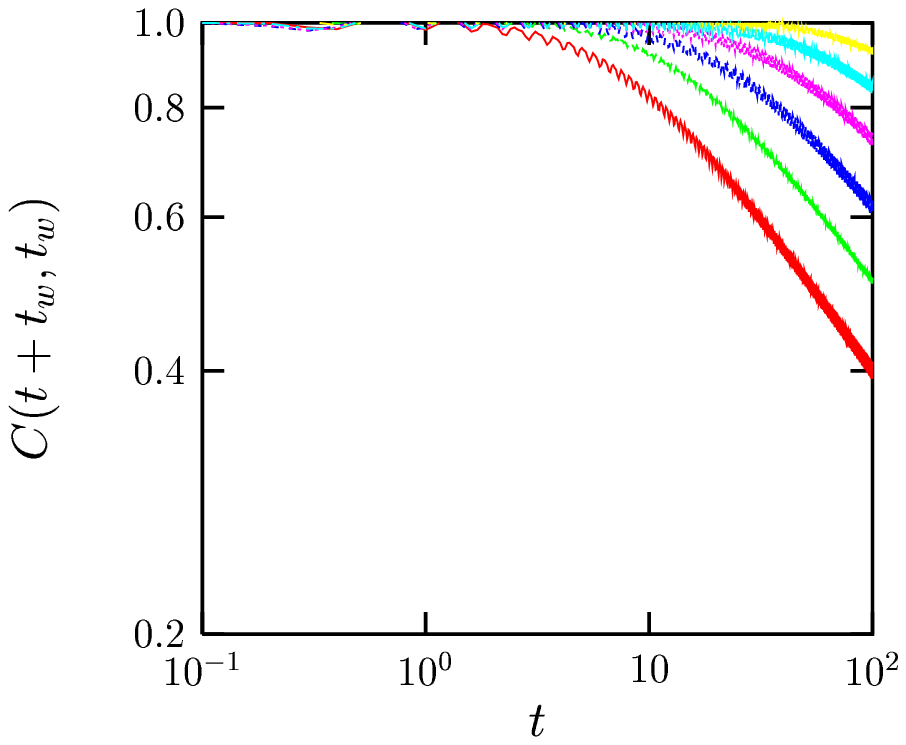}
}
\caption{The auto-correlation function $C_{k=0}(t+t_w,t_w)$ for
several choices of the waiting times $t_w$. Clockwise from the 
top-left panel:  
the relaxing spherical $p$ spin model at $T=0.2$, its quantum extension
at $T=0$, 
the vibrated case  at $T=0$ and the ``sheared'' one at $T=0.2$ 
(aging is interrupted in this case).}
\label{figC}
\end{figure}

Below the glass transition the decay after a waiting-time $t_w$ 
also occurs in two steps. There is a first stationary decay toward 
a plateau whose value depends on the external parameters $(T$, field, etc.)
and $k$, and a second slow decay toward zero that occurs in a waiting-time 
dependent scale $t_\alpha(t_w,T)$, with $t_\alpha(t_w,T)$ 
a growing function of $t_w$: the older the system the slower the decay.
The top-left panel in Fig.~\ref{figC} displays the decay 
of the self correlation $C(t+t_w,t_w)$, 
for instance $C(t+t_w,t_w)=
N^{-1} \sum_i \langle s_i(t+t_w) s_i(t_w) \rangle$, with $s_i$ 
the dynamic variables and the brackets representing an average over different
thermal histories, in the spin system).

The two step relaxation can be simply visualized in a number of 
examples that provide different interpretations for the glassy
dynamics. 
Phenomenological descriptions of 
glassy dynamics based on the motion of a point, that represents
the full system, in a roughed free-energy landscape are popular
(more precisely, an average over subsystems is needed to obtain
continuous results).
However, since phase space is infinite dimensional ($2^N$ dimensional 
in a spin system) the intuition based on the knowledge of diffusion 
processes in finite dimensional spaces are misleading and can lead to 
misconceptions when applied to the glassy problem. The free-energy 
landscape ({\sc tap} free-energy), the non-equilibrium dynamics,  
and the relation between them, are completely known for  
solvable glassy models like the $p$ spin model. In the asymptotic
limit, the point that represents the system in phase space approaches a 
path made of flat directions, the {\it threshold}, 
that is higher in free-energy density than the equilibrium level~\cite{Cuku}.
The rapid first step
relaxation is related to the rapid motion ``transverse'' to the flat 
directions while the slow $\alpha$-relaxation has to do with the slow diffusion
along the path and it is not an activated process over barriers.  
Since phase space is infinite dimensional, times that diverge 
with $N$ are needed to penetrate 
below the threshold. These processes are now of a different type, 
being due to activation from one well in the 
{\sc tap} free energy to another. When contact is made with real materials
such long times might be unrealistic, unless close to $T_g$, where 
transitions between metastable states are certainly 
easier 
and can be responsible for 
the observed cooling rate dependences. (With a slower annealing the 
system spends longer periods at higher temperatures where transitions 
are more favorable and it can then penetrate more deeply below the threshold.
All (one-time) quantities such as the internal energy density 
will consequently approach lower values and be closer to the ``optimum''.) 
The relevance of saddles in the free-energy landscape is not peculiar to 
the non equilibrium dynamics of mean-field models~\cite{landscape}.  

The two step relaxation of the self-correlation 
toward and away from a plateau 
captures the well-know {\it cage effect}. In a system of interacting 
particles, any particle is surrounded 
by a small number of neighbors that form a cage around it. 
The short time-difference dynamics corresponds to the rapid rattling of 
particles in their cages and it resembles the dynamics of a
system in equilibrium in that it 
is stationary. 
The second step is instead due to the 
destruction of the cages, each particle escapes its own cage and there is 
a structural rearrangement.
How much can the particles decorrelate within the cages depends on 
temperature: the height of the  
plateau in the correlation, that is related to the ``size'' of the cage, 
decreases with increasing temperature.
The time needed to destroy the cages, $t_\alpha(t_w, T)$,
depends on the waiting-time (and $T$). Even if in these models
there is no notion of interacting particles, the models capture 
a cage effect and they predict a form 
for the decay of $C_k(t+t_w,t_w)$ that has been later observed 
numerically in more realistic glassy models~\cite{numerics}.

Certainly, such a two-step process is also captured by coarsening models,
the simplest examples being spin models undergoing domain growth. In these,
after the quench from high temperatures, two ordered phases grow 
with none of them conquering the sample if the thermodynamic limit has 
been taken at the outset. However, it has remained an open question if 
any such growing order exists in most glassy systems. It  is also 
worth mentioning that models with an infinite hierarchy of relaxation 
times exist~\cite{Cuku2,space}.
 
The two-step decay is robust with respect to changes in the microscopic 
dynamics. On the one hand, the same kind of two-step relaxation 
is observed using molecular dynamics or Montecarlo to simulate the 
dynamics of a given model. On the other hand, more important changes in the  
driving  dynamics slightly modify this picture, as shwon below.

The simplest effect of quantum fluctuations is to introduce oscillations
in the first step of relaxation. These disappear at long enough 
time-differences and they are totally suppressed from the second decay,
that superficially looks classical~\cite{Culo,quantum_others}. 
(A more dramatic effect of 
quantum mechanics is related to the very strong role 
played by the quantum environment 
on the dynamics of a quantum system. Indeed, the location of the 
transition line strongly depends on the type of quantum 
bath considered and on the strength of the  coupling between system and 
environment~\cite{quantum_bath}.) The top-right panel in Fig.~\ref{figC} 
shows the decay of the 
symmetrical correlation for the quantum extension of a $p$ spin model
in its glassy phase.

A weak shear has a spectacular effect on the relaxation of macroscopic
correlations. It introduces a shear-dependent time scale $t_{\sc sh}$ 
and, for waiting-times that are longer than $t_{\sc sh}$, aging is
interrupted. The effective age is a decreasing function of the 
pumped energy. This effect has been known for long 
experimentally~\cite{rheology, Onuki}  and it is related to the 
shear-thinning behavior in Ref.~\cite{rheology}. 
It has been found and explored recently within the theoretical framework
that we review~\cite{jorge,rheology_theor}. Under  a weak shear
the decay still occurs in two steps though it is fully stationary 
(for waiting times that are longer than $t_{\sc sh}$), as shown in 
in the bottom-left panel in Fig.~\ref{figC}. 

The effect of an ac-field on a glassy system is complex
and it is somehow similar to the action  of a quantum bath on 
a quantum glass . 
When the test glassy model is externally perturbed by an oscillatory 
field, $H \sin(\omega t)$, the correlation oscillates 
with a frequency that is determined by the one of the external drive. If one 
explores the dynamics by using stroboscopic time, that is to say by using a 
single point for each  cycle, the relaxation becomes monotonic and 
smooth with a clearcut time-separation. Aging is not interrupted by the 
external perturbation but the location of the 
liquid-to-glass transition in the $(T,H,\omega)$ parameter space 
depends on $\omega$. The effect of an oscillatory field on a spin system 
is weaker than the one of a 
dc field with the same amplitude: 
the projection of the glassy phase in the $(T,H,\omega)$ parameter space 
on to the $(T,H)$ plane, is larger for an ac-field than 
for a dc one~\cite{granular_us}. The bottom-right panel in 
Fig.~\ref{figC} displays the decay of the 
self-correlation for different waiting times when an ac field is applied. 
Temperature is zero and there is no first decay since the plateau is
at $C=1$. Waiting-time dependences are apparent.  

The threshold level in the {\sc tap} free-energy still plays an important 
role in all these modified problems. A quantum {\sc tap} free-energy density 
can be defined and, for $p$ spin-like models, it is similar to 
the classical one with a threshold level, many minima below it, 
etc~\cite{Bicu}. 
In systems perturbed by non-potential forces, 
the energy injection serves to keep the system in a 
steady state above the threshold level, the higher the stronger the drive. 
 
\section{Responses and effective temperatures}

The solution to dynamics of glassy models
relates the spontaneous fluctuations 
$C_k(t+t_w,t_w)$ and the induced ones, $\chi_k(t+t_w,t_w)$, defined as 
the integrated linear
response over a finite time interval $[t_w,t+t_w]$. 
It was shown in \cite{Cukupe} that the slope of the plot 
of $\chi_k(t+t_w,t_w)$ against $C_k(t+t_w,t_w)$ for $t_w$ fixed and using 
$t$ as a parameter, defines a {\it measurable} 
effective temperature.

\begin{figure}[t]
\centerline{
\epsfxsize=3in
\epsffile{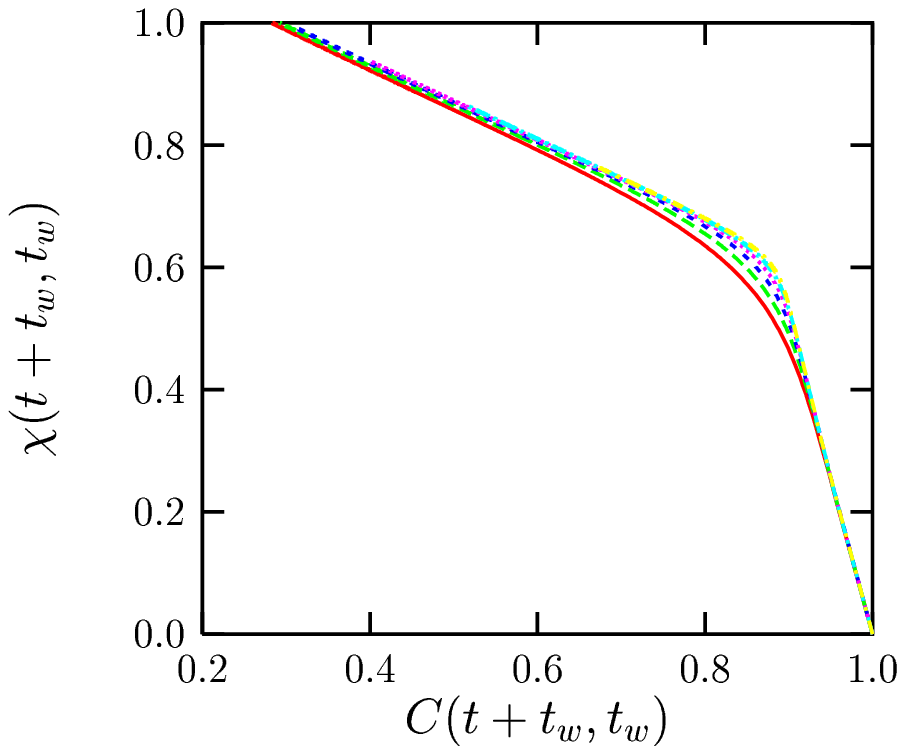}
\hspace{.25cm}
\epsfxsize=3in
\epsffile{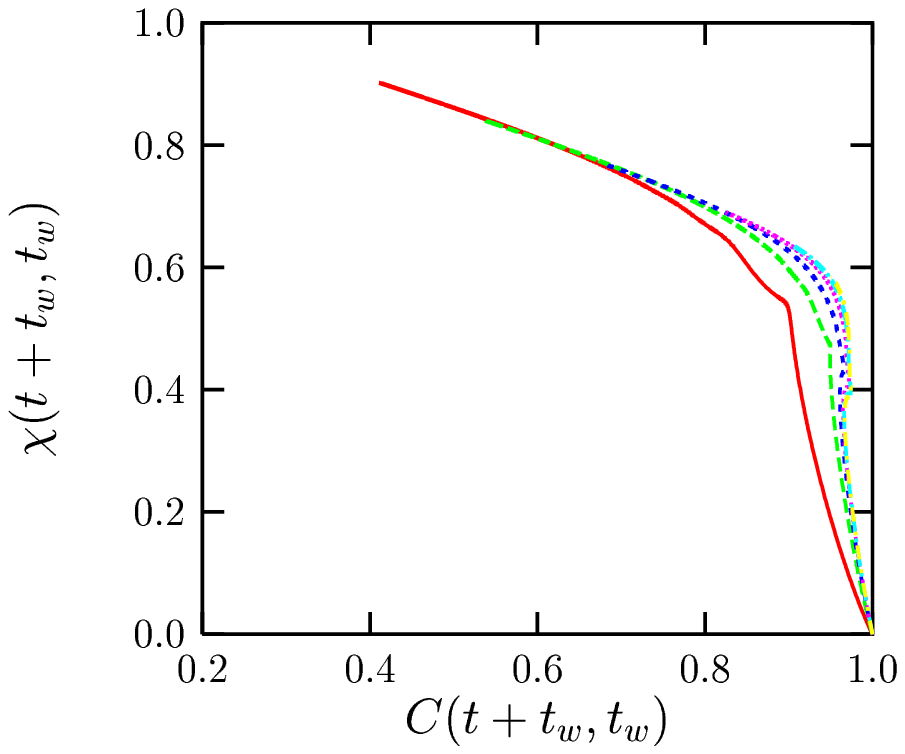}
}
\centerline{
\hspace{.7cm}
\epsfxsize=3in
\epsffile{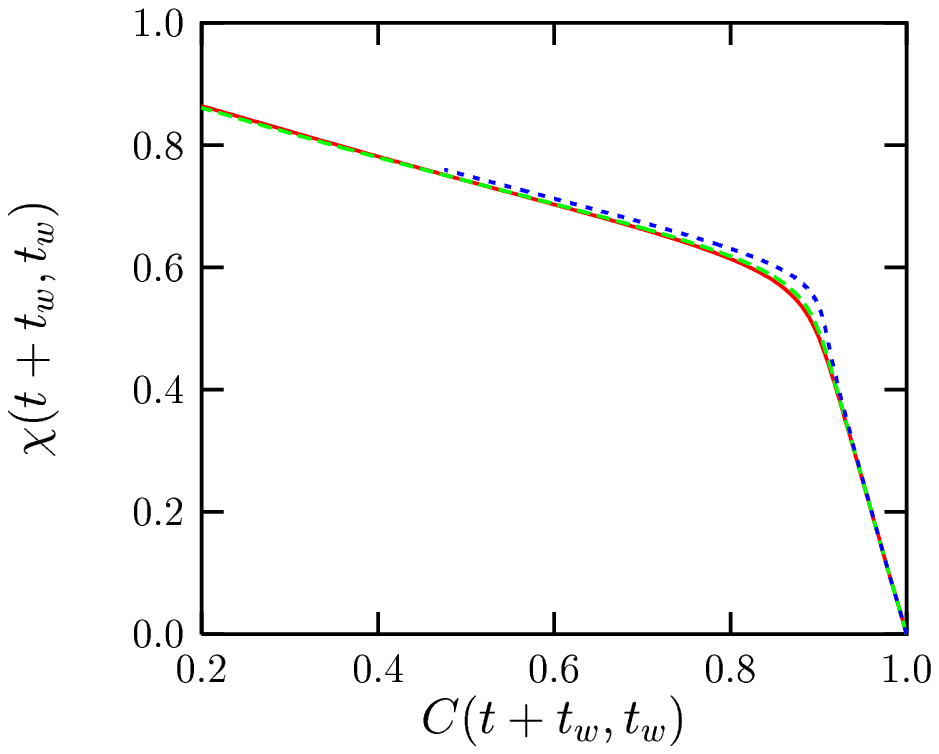}
\hspace{-1cm}
\epsfxsize=3.12in
\epsffile{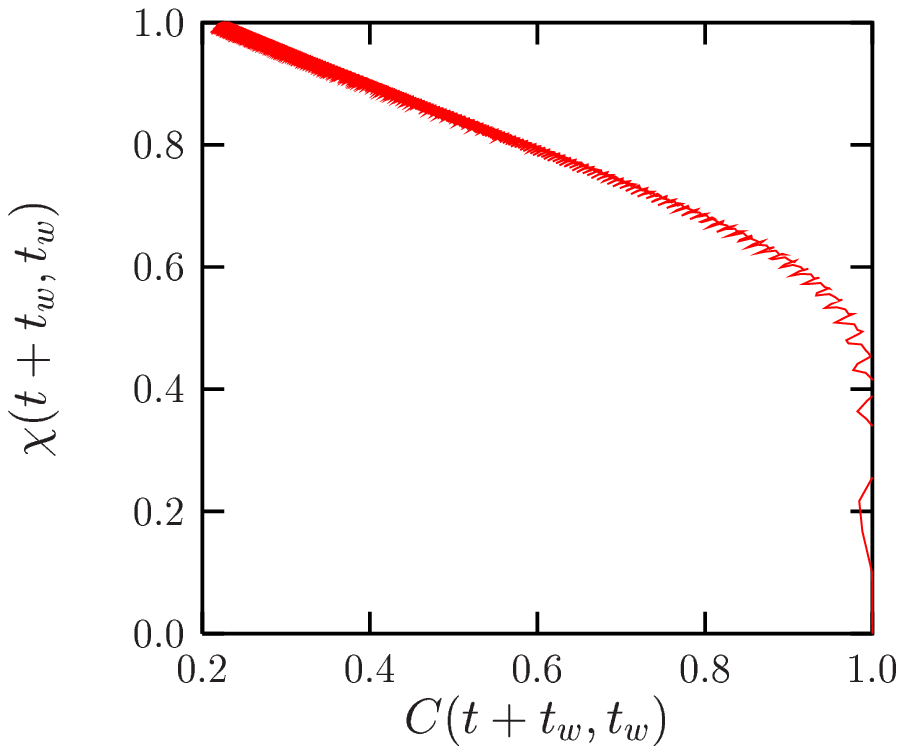}
}
\caption{The parametric plot of the integrated response, $\chi$, against
the auto-correlation, $C=C_{k=0}(t+t_w,t_w)$, for several values 
of the waiting time. Clockwise: the relaxing spherical $p$ spin model at 
$T=0.2$, its quantum extension at $T=0$, 
the vibrated case at $T=0$ and the ``sheared'' one at $T=0.2$. }
\label{figchi}
\end{figure}

In classical thermal systems without external forces 
the effective temperature takes different values in 
different time-scales~\cite{Cuku}. 
In the rapid time-scale, 
where $C_k$ decays from $1$ to the plateau $q_k$,  
the equilibrium fluctuation-dissipation  theorem ({\sc fdt})
is satisfied and the effective temperature is that of the 
environment. The parametric construction yields a straight line of 
slope $-1/T$. This result is natural. For a particle system, this 
regime corresponds to the motion of particles within their  cages 
that strongly resembles equilibrium. Consequently, the kinetic energy in a 
glass satisfies equipartition with the temperature of the environment. 
In a coarsening spin system it corresponds to the 
thermal spin flips within the domains that is also very reminiscent of an 
equilibrium problem. During the $\alpha$ structural relaxation instead  
the effective temperature takes a different value, that is close (but not 
identical) to the critical temperature $T_d$ (see the top-left panel in 
Fig.~\ref{figchi}). In $p$ spin like models 
there is only one independent correlation and response. In the extensions
based on models of finite dimensional manifolds in random potentials
all the {\sc fd} relations that one can construct 
using different $k$ lead to the same value of the effective
temperature, $T_{\sc eff}(k) = T_{\sc eff}$~\cite{space,Arnulf}. 
The reason for this 
property is that all observables evolving in the same time-scale,
and interacting, 
equilibrate and hence acquire the same value of the
effective temperature~\cite{Cukupe}. 
This property has also been checked numerically in Lennard Jones 
mixtures~\cite{Ludo-unpub}.
(In models with a hierarchy of 
time-scales, each time-scale has its own effective temperature~\cite{Cuku2}.)

The quantum {\sc fdt} is a retarded 
non-linear relation between the spontaneous symmetrical 
correlations and the linear response. Quantum glassy 
systems have (at least) two time-regimes~\cite{Culo,quantum_others}. 
In the rapid regime 
the quantum {\sc fdt} holds. This is the reason why the plot 
on the right-top panel in Fig.~\ref{figchi} has a strange form and 
does not show a straight line when the correlation varies from 
$1$ to $q_k$. Instead, in the second time-regime 
the parametric construction  yields a
linear relation between $\chi_k$ and $C_k$. 
The implications 
of this result are very interesting. The classical 
{\sc fd} relation develops, with a slope
given by a non-trivial $T_{\sc eff}$ (even at zero external temperature). 
The dynamics looks classical and a decoherent effect appeared. 

In systems perturbed by non-potential forces, like shear, a non-equilibrium 
steady state is reached that is characterized by two time-scales with 
different effective temperatures. The $p$ spin models with asymmetric 
couplings~\cite{Cukulepe} 
is the mean-field  realization of a super-cooled liquid or a 
glass perturbed in this way~\cite{rheology_theor} (see the bottom-left panel 
in Fig.~\ref{figchi}). 
Binary Lennard-Jones mixtures show the same 
phenomenology~\cite{JL-Ludo}.  

In a granular system the microscopic dynamics is not thermal; 
the dynamics is driven by the energy pump and dissipation is not 
proportional to the velocity.
If two time-regimes develop, the fast regime should not 
be characterized by an effective temperature. Instead, 
if the structural relaxation is slow and similar
to that of a glassy system, there should be an effective temperature
controlling this slow time-scale, in the same way as in the 
quantum system we recovered a classical effective temperature in this 
regime. This scenario has been checked using a 
$p$ spin model perturbed by an ac force (where the bath temperature is 
still relevant, right-bottom panel in Fig.~\ref{figchi})~\cite{granular_us} 
and more realistic models 
of granular matter externally sheared \cite{Makse}. In the 
vibrated case, the existence of an effective temperature is 
better visualized in stroboscopic time, when one divides away
the oscillations directly related to the external drive. 

\section{Conclusions}

We have exhibited several features of the dynamics of 
simple models that have also appeared in more realistic ones and real 
systems. Without wanting to stretch 
the domain of applicability of 
these models too far (some of their limitations are obvious!),  
we believe that the scenario they suggest is very close 
to the one of the dynamics of generic systems in the limit of small 
entropy production (glassy, gently driven).

The number of models showing relaxations as in Fig.~\ref{figC}
and the existence of an effective temperature 
is now very large (spin systems with and without disorder,
models of interacting particles and  
polymers, 
kinetically constrained models, realistic models 
of granular matter, etc.). Moreover, in 
Figs.~\ref{figC}  and \ref{figchi} we have shown the effect of 
three rather different 
ways of modifying the microscopic dynamics that still leave 
untouched the separation of time scales and the existence of a non-trivial 
effective temperature for the slow one (apart from slightly 
changing its precise value). 
One may wonder why the dynamic features presented here are so robust. 

One reason for this is technical.
For any dynamical 
problem one can write Schwinger-Dyson equations for the evolution 
of the two-point correlations and the corresponding linear responses. 
These equations involve vertices and self-energies 
that depend on the same two-point
functions but that, in general, 
cannot be calculated explicitly. 
Two ways of approximating these equations are~\cite{Mode}:
\newline
-- to change the departing Hamiltonian to a mean-field like one
for which explicit Schwinger-Dyson equations are exactly derived. 
\newline
-- to approximate the vertices and self-energy of the original theory 
with a self-consistent procedure (MCT~\cite{Arnulf}, SCSA, etc) in order
 to render them explicit.
The structure of the approximate equations is very constrained 
by the symmetries in the problem (including the rather abstract
supersymmetry of stochastic processes) that act as a guideline to
derive them.  

As explained in \cite{Mode}, 
there is a one-to-one correspondence between one method and the 
other. It is then no surprise that the behavior of 
mean-field models, approximation schemes and real
systems are indeed very similar. 

The general Schwinger-Dyson equations have a structure such that, 
when a fast time-scale can be separated from the 
slow one(s), the existence of a non-trivial effective temperature
corresponds to a symmetry breaking~\cite{Cuku-japan}. The structure we have 
shown here, even if initially derived in the mean-field $p$ spin model,
is then consistent even for finite dimensional problems (it does not 
include non-perturbative effects though). This argument 
explains why it is easier to change the time-dependences in the 
relaxation (e.g. from aging to stationary in the sheared case) than
the organization of effective temperatures~\cite{note}. However, whether
one given system will realize this many temperature structure 
or not cannot be easily guessed from 
looking at the microscopic Hamiltonian (much as it is difficult to guess the 
order of a static phase transition for a model with complicated 
interactions).

\end{document}